\documentclass[fleqn,twoside]{article}
\usepackage{espcrc2}

\usepackage{graphicx}
\usepackage[figuresright]{rotating}

\newcommand{\AmS}{{\protect\the\textfont2
  A\kern-.1667em\lower.5ex\hbox{M}\kern-.125emS}}

\title{Phase transitions in large $N_c$ heavy quark potentials}

\author{F. Bigazzi\address{Physique Th\'eorique et Math\'ematique and International Solvay
Institutes,\\ Universit\'e Libre de Bruxelles; CP 231, B-1050 Bruxelles, Belgium.}\thanks{Work supported by the EU contract MRTN-CT-2004-005104, the FWO -
Vlaanderen project G.0235.05, the Federal Office for
STC Affairs programme P6/11-P, the Belgian
FRFC (G.2.4655.07), IISN (G.4.4505.86), the IAPP (Belgian Science Policy), a NWO VIDI grant 016.069.313 and
INTAS contract 03-51-6346.
Preprint numbers: KUL-TF-08/28, ITP-UU-08/64, SPIN-08/51.
We thank C. N\'u\~nez for collaboration in the research reported here.
\emph{F.B. and A.L.C. would like to thank the italian students, parents and scientists for their activity in support of public education and
research.}}, A. L. Cotrone\address{Institute for theoretical physics, K.U. Leuven;
Celestijnenlaan 200D, B-3001 Leuven,
Belgium.} and A. Paredes\address{Institute for Theoretical Physics, Utrecht University; Leuvenlaan 4,
3584 CE Utrecht, The Netherlands.}}

\begin{document}

\begin{abstract}
We describe phase transitions in the heavy quark potential in planar gauge theories having wrapped D5-brane string duals.
A new phase transition, previously unnoticed in these models,
is driven by the source of a large dimension operator.
Another transition, which has already been described in a previous paper, is driven by the presence of light dynamical flavors.
Both transitions connect a Coulomb-like phase to a confining linear phase.

\vspace{1pc}
\end{abstract}

\maketitle
\setcounter{footnote}{0}

\section{STRINGS AND GAUGE THEORIES}

String theory is a valuable mean of studying strongly coupled gauge theories,
allowing to extract important qualitative and quantitative informations for QCD.
In this note we will describe phase transitions in the heavy quark potential, induced either by the source of a higher dimensional operator, or by the presence of light dynamical flavors (in this case, the transition already appeared in \cite{transition}), by means of the string description of certain planar gauge
theories.
The theories are at zero temperature, so these effects are examples of quantum phase transitions.
Their presence in the unflavored case\footnote{In this case, the string configuration used to calculate the heavy quark potential is stable rather than metastable.} indicates that they are not induced by the ``smearing technique'' used in \cite{transition} to derive the gravity solution in presence of flavors.
Since the strongly coupled regime of theories with large dimension operators or light quarks is extremely difficult to study with any other technique (e.g. the lattice \cite{Bali:2005fu}), the present  results provide novel information on its physics.

\section{DESCRIPTION OF THE THEORIES}

Supergravity solutions dual to theories with dynamical flavors can be conveniently derived by homogeneously smearing in the internal space the branes corresponding to the flavor degrees of freedom \cite{Bigazzi:2005md}.
Following this procedure, the authors of \cite{Casero:2006pt} derived a family of solutions dual to flavored versions of a confining theory corresponding to wrapped D5-branes \cite{Maldacena:2000yy}.
In their solutions the flavors are massless.
We are interested in adding mass to the dynamical quarks.
A simple way of achieving this is to glue unflavored solutions (also contained in \cite{Casero:2006pt}) to the massless-flavored ones, at a radial position which is dual to the quark mass.
This procedure reflects the physical situation where the effects of the massive flavors are vanishing at energies smaller than their mass, while they are present at larger energies.
From the Born-Infeld point of view the construction is an approximation, but from the experience with other models where one can compare it with the exact solution \cite{paper2}, it is expected to correctly capture the main qualitative features of the physics.

\begin{figure}[htbp] 
 \includegraphics[width=0.234\textwidth]{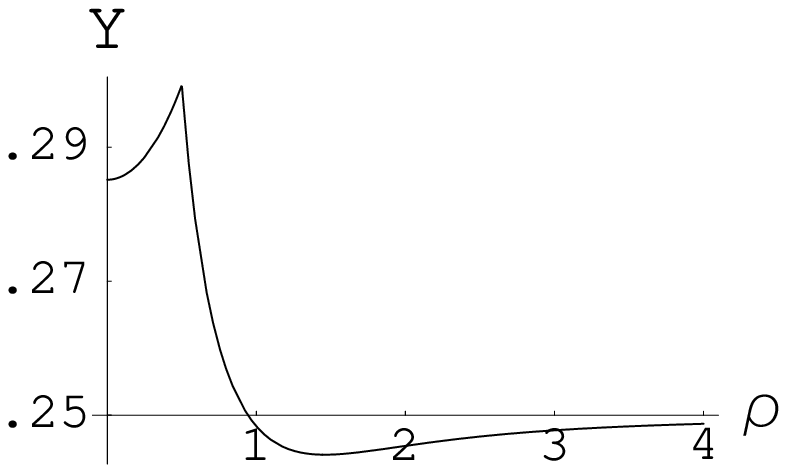}\includegraphics[width=0.234\textwidth]{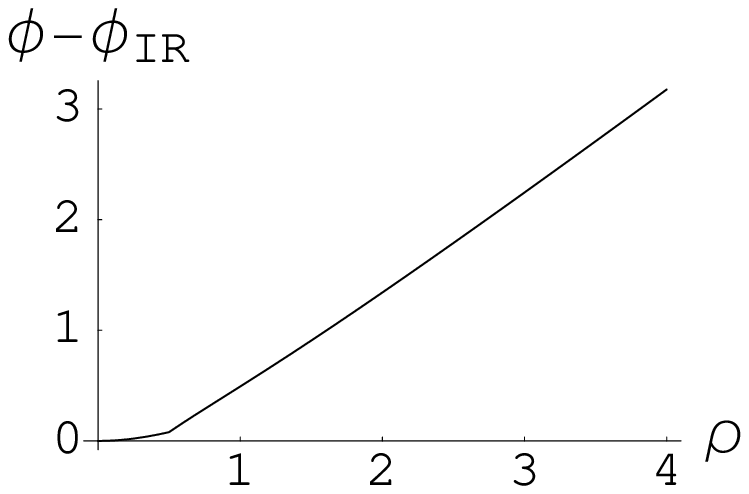}
 \caption{The typical form of the $Y,\phi$ functions.}
 \label{x1rhoM1}
 \end{figure}

The relevant part of the metric reads
\begin{eqnarray} ds^2 &=& \alpha'e^\phi N_c \Big[ \frac{1}{\alpha'g_sN_c}dx_{1,3}^2 + 4 Y d\rho^2 + ... \Big] \nonumber
\label{SDmetric} \end{eqnarray}
and a typical solution 
is shown in figure \ref{x1rhoM1}.

If continued up to the UV,
the unflavored solutions describe a deformation of the solution in \cite{Maldacena:2000yy} by the source of a higher dimension operator \cite{Casero:2006pt,HoyosBadajoz:2008fw}.
Its presence drastically modifies the large $\rho$ behavior of the geometry and the dilaton is not linear anymore.
Notice, though, that the flavored solution described above is constructed in such a way that
the source of this operator is kept to zero.
In fact, the asymptotics are qualitatively the same as in the undeformed solution \cite{Maldacena:2000yy}.

\section{PHASE TRANSITIONS}
\begin{figure}[htb]
\includegraphics*[scale=0.75]{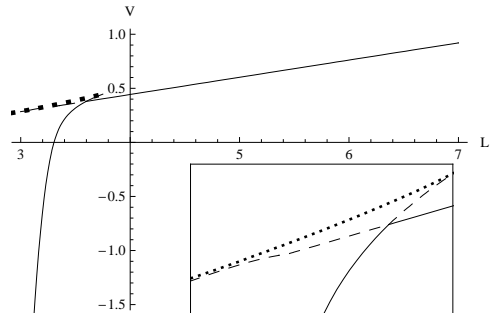}
\caption{Heavy quark potential for small flavor mass, with a zoom of the critical region. The plot for the case of large operator source is analogous.}
\label{gloriousturnaround}
\end{figure}

The potential $V(L)$ between two heavy external quarks at distance $L$ can be computed by means of a classical string in the backgrounds described above \cite{Rey:1998ik}.
The presence of the dynamical light flavors (or of the higher dimension operator) has dramatic effects.
When the mass of the flavors is above a certain critical value (or the source of the operator is below a certain critical value), the potential\footnote{With dynamical flavors $q$, a bound state of heavy quarks $\bar Q, Q$ is metastable due to the decay $\bar Q Q\rightarrow \bar Q q+ \bar q Q$. Our potential refers only to the metastable configuration.} is qualitatively similar to the Cornell one \cite{cornell}.
But when the mass of the dynamical flavors is smaller than the critical value (or the source of the operator is larger than the critical value), the heavy quark potential displays a phase transition between the Coulomb-like phase at small distance $L$ and the confining linear phase at large $L$.
The phase transition corresponds to a sudden decrease of the local string tension, reflected in the discontinuity of the slope in the plot in figure \ref{gloriousturnaround}.\footnote{Phase transitions of similar
nature induced by different physical situations were
first reported in \cite{Brandhuber:1999jr}.}
This effect of course vanishes in the flavorless limit (or in the source-less limit for the higher dimension operator).

\end{document}